\documentclass[prb,twocolumn,amsmath,amssymb,showpacs]{revtex4}

\usepackage{graphicx}

\begin{document}

\newcommand{\etal}{{\it et al.}\/} 
\newcommand{\gtwid}{\mathrel{\raise.3ex\hbox{$>$\kern-.75em\lower1ex\hbox{$\sim$}}}} 
\newcommand{\ltwid}{\mathrel{\raise.3ex\hbox{$<$\kern-.75em\lower1ex\hbox{$\sim$}}}}

\title{Origin of Gap Anisotropy in Spin Fluctuation Models of the Fe-pnictides}

\author{T.A.~Maier} \email{maierta@ornl.gov} \affiliation{Center for Nanophase Materials Sciences and Computer Science and Mathematics Division, Oak Ridge National Laboratory, Oak Ridge, TN 37831-6494, U.S.A.} 

\author{S.~Graser} \email{graser@phys.ufl.edu} \affiliation{Center for Electronic Correlations and Magnetism, Institute of Physics, University of Augsburg, D-86135 Augsburg, Germany}

\author{D.J.~Scalapino}\email{djs@physics.ucsb.edu} \affiliation{Department of Physics, University of California, Santa Barbara, CA 93106-9530, U.S.A.}

\author{P.J.~Hirschfeld} \email{pjh@phys.ufl.edu} \affiliation{Department of Physics, University of Florida, Gainesville, FL 32611, U.S.A.}

\date{\today} 

\begin{abstract}
	We discuss the large gap anisotropy found for the $A_{1g}$ ($s$-wave) state in RPA spin-fluctuation and functional renormalization group calculations and show how the simple arguments leading to isotropic sign-switched s-wave states in these systems need to be supplemented by a consideration of pair scattering within Fermi surface sheets and between the individual electron sheets as well. In addition, accounting for the orbital makeup of the states on the Fermi surface is found to be crucial. 
\end{abstract}

\maketitle


With one or two exceptions, the newly discovered Fe-pnictide superconductors are created by doping parent materials which manifest a magnetically ordered ground state. Since critical temperatures are as high as 56K, and experiments have shown enhanced magnetic response relative to ab initio electronic structure calculations, it is natural to consider spin fluctuation type pairing models for these systems. There have been several calculations of this general type, which have reported gaps with significant anisotropy on the multisheeted Fermi surface of these materials \cite{kuroki:prl08,kuroki:prl09,kuroki:njp09,qi:arxiv08,barzykin:arxiv08,bang:prb08,yao:arxiv08,sknepnek:prb09,wang:prl09,chubukov:prb08}. For the most part, a ground state with A1g symmetry has been identified, with a nearby $d_{x^2-y^2}$ (in the unfolded Brillouin zone) pairing eigenvalue. These calculations involve fairly realistic representations of the electronic structure near the Fermi surface, and are sufficiently complicated that the precise physical effects leading, e.g. to a particular pairing channel or to gap anisotropy can be obscured.

At the same time, there is a simple argument originally given by Mazin {\it et al.} \cite{mazin:prl08} which suggests that the order parameter in the system should have a sign-switched s-wave structure \cite{mazin:prl08,chubukov:prb08,yanagi:jpsp08,sknepnek:prb09,chubukov:arxiv09}. It is based on the argument that the scattering of pairs between the $\alpha$-hole Fermi surfaces around the $\Gamma$ point and the $\beta$-electron Fermi surfaces around the $X(\pi,0)$ and $Y(0,\pi)$ points of the unfolded Brillouin zone (see Fig.~1) is dominant in the system due to the near-nesting of the small sheets. In this picture, the system can maximize its condensation energy by forming isotropic order parameters but switching sign between the $\alpha$ and $\beta$ sheets to take advantage of the interband pair scattering. This change in sign has the added benefit of reducing the short range on site Coulomb repulsion.\cite{mazin:arxiv09} It is therefore at first sight surprising to find that RPA spin-fluctuation calculations\cite{kuroki:prl08,graser:njp09} as well as functional renormalization group studies\cite{wang:prl09} find a highly anisotropic $A_{1g}$ $s$-wave gap. Here we investigate this, as well as explore the question of what the anisotropy can tell us about the pairing interaction. One answer that has been given is that the momentum dependence of the fluctuation-exchange pairing interaction can drive the anisotropy \cite{graser:njp09}. However, we will see that the orbital make-up of the states on the Fermi surface and the suppression of the short range Coulomb interaction \cite{chubukov:arxiv09} also play key roles in favoring an anisotropic gap.

\begin{figure}
	[htbp] 
	\includegraphics[width=7cm,clip,angle=0]{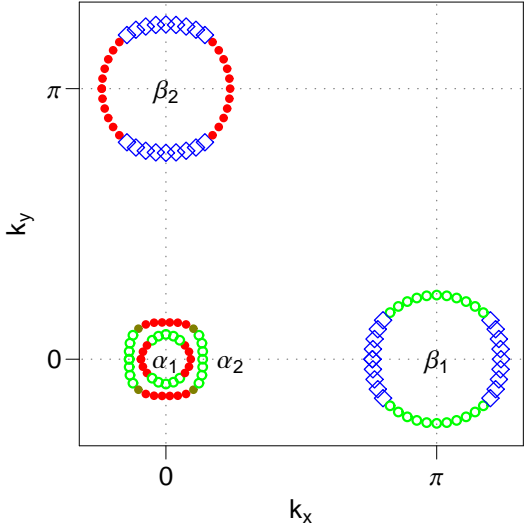} \caption{(color online) The Fermi surface of the 5-orbital tight-binding model \cite{graser:njp09}. The main orbital contributions are shown by the following colors/symbols: $d_{xz}$ (red/solid circles), $d_{yz}$ (green/open circles), $d_{xy}$ (blue/diamonds).} \label{fig:1} 
\end{figure}

In the following we examine the pairing strengths for processes which involve a pair of electrons scattering on and between the four Fermi surfaces shown in Fig.~\ref{fig:1}. 
These were calculated using a 5-orbital ($d_{xz}$, $d_{yz}$, $d_{xy}$, $d_{x^2-y^2}$, $d_{3z^2-r^2}$) tight-binding fit to the DFT bandstructure calculations of Cao et al \cite{cao:prb08}. The tight-binding parameters for an orbital basis that is aligned parallel to the nearest neighbor Fe-Fe direction are given in the appendix of Ref.~\onlinecite{graser:njp09}. Here we will label the two hole Fermi surfaces $\alpha_1$ and $\alpha_2$ and the two electron Fermi surfaces $\beta_1$ and $\beta_2$ as indicated in Fig.~\ref{fig:1}.

\begin{figure}
	[htbp] 
	\includegraphics[width=9cm,clip,angle=0]{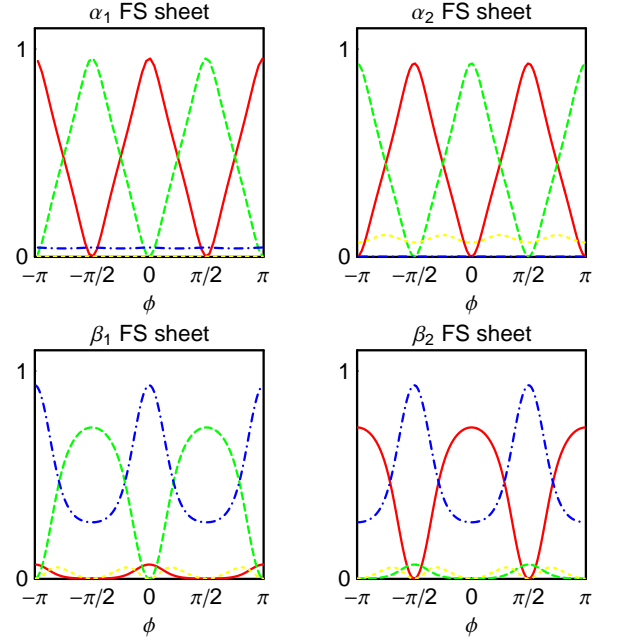} \caption{(color online) The orbital weights as a function of the winding angle $\phi$ on the different Fermi surface sheets \cite{graser:njp09}. The different colors/lines refer to $d_{xz}$ (red/solid lines), $d_{yz}$ (green/dashed lines), $d_{xy}$ (blue/dash-dotted line), $d_{x^2-y^2}$ (yellow/short-dashed line).} \label{fig:2} 
\end{figure}

The orbital weights $a^t_i(k)$ of the states on the various Fermi surfaces are shown in Fig.~\ref{fig:2} and illustrated by the colors in Fig.~\ref{fig:1}. 
Here $i$ designates the Fermi sheet $\alpha_1$, $\alpha_2$, $\beta_1$, and $\beta_2$ and $t$ the orbital ($d_{xz}$, $d_{yz}$,\dots). The dominant orbitals contributing to the $\alpha_1$ and $\alpha_2$ sheets are the $d_{xz}$ and $d_{yz}$ orbitals. The upper and lower parts of the $\alpha_1$ sheet are $d_{yz}$-like and the left and right hand sides have $d_{xz}$ character. The opposite behavior is seen on the $\alpha_2$ sheet. On the $\beta_1$ sheet, the upper and lower surfaces have dominantly $d_{yz}$ character while the ends along the $k_x$-axis have $d_{xy}$ character. Similarly, the $\beta_2$-sheet is made up of $d_{xz}$ along the sides and $d_{xy}$ on the $k_y$-axis ends \cite{mazin:prl08,graser:njp09,lee:prb08}.

Now consider the scattering of a pair ($k'\uparrow,-k'\downarrow$) on the $\alpha_1$ Fermi surface to a pair ($k\uparrow,-k\downarrow$) on the $\beta_1$ Fermi surface. The strength of this scattering depends upon 
\begin{eqnarray}
	\Gamma_{ij}(k,k')=&&\\
	&&\hspace{-1.5cm}\sum_{stpq}a^{t,*}_{\nu_i}(-k)a^{s,*}_{\nu_i}(k){\rm Re} \left[\Gamma^{pq}_{st}(k,k',0)\right]a^p_{\nu_j}(k')a^q_{\nu_j}(-k')\nonumber \label{eq:1} 
\end{eqnarray}
and involves orbital weight factors and, within the fluctuation exchange approximation the orbital dependent vertex 
\begin{eqnarray}
	\Gamma^{pq}_{st}(k,k')=&&\\
	&&\hspace{-1.5cm}\left[\frac{3}{2}\left(U^s\chi^{\rm RPA}_s U^s+\frac{U^s}{2}\right)-\frac{1}{2}\left(U^c\chi^{\rm RPA}_c U^c-\frac{U^c}{2}\right)\right]^{tq}_{ps}\nonumber \label{eq:1a} 
\end{eqnarray}
Here, the momenta $k$ and $k'$ are restricted to the different Fermi surface sheets $C_i$ with $k\in C_i$ and $k'\in C_j$. The interaction matrices $U^s$ and $U^c$ contain the onsite intra- and inter-Coulomb interactions along with the exchange couplings and $\chi^{\rm RPA}_s$ and $\chi^{\rm RPA}_c$ are the RPA spin and charge orbital susceptibilities. We will use numerical results obtained from earlier work \footnote{Here we have set the onsite intra-orbital $U=5/3$, the exchange $J=2J'=U/8$ and the inter-orbital $V=U-3/4J-J'$. These represent typical parameters for which the leading pairing strength occurs in the singlet $A_{1g}$ state. We find similar behavior for a range of parameters.} but our results are not dependent on the precise values of the parameters. Rather, what is important to note is that the dominant pairing interaction is found to arise from the spin-fluctuation term $\frac{3}{2}U^s\chi^{\rm RPA}_sU^s$ and the short range Coulomb contact interactions $\frac{3}{4}U^s$ and $\frac{U^c}{4}$ oppose the pairing. As noted by Mazin and Schmalian \cite{mazin:arxiv09}, for a simple two-band model with equal density of states, the short range Coulomb repulsion vanishes for a sign switched $s$-wave gap. However, as they point out, when there is an asymmetry in the density of states, some fraction of the Coulomb-repulsion remains. We will see that it is both the orbital weight factors and the further suppression of the Coulomb interaction that lead to the anisotropy of the gap in the realistic case.

\begin{figure}
	[htbp] 
	\includegraphics[width=8cm,clip,angle=0]{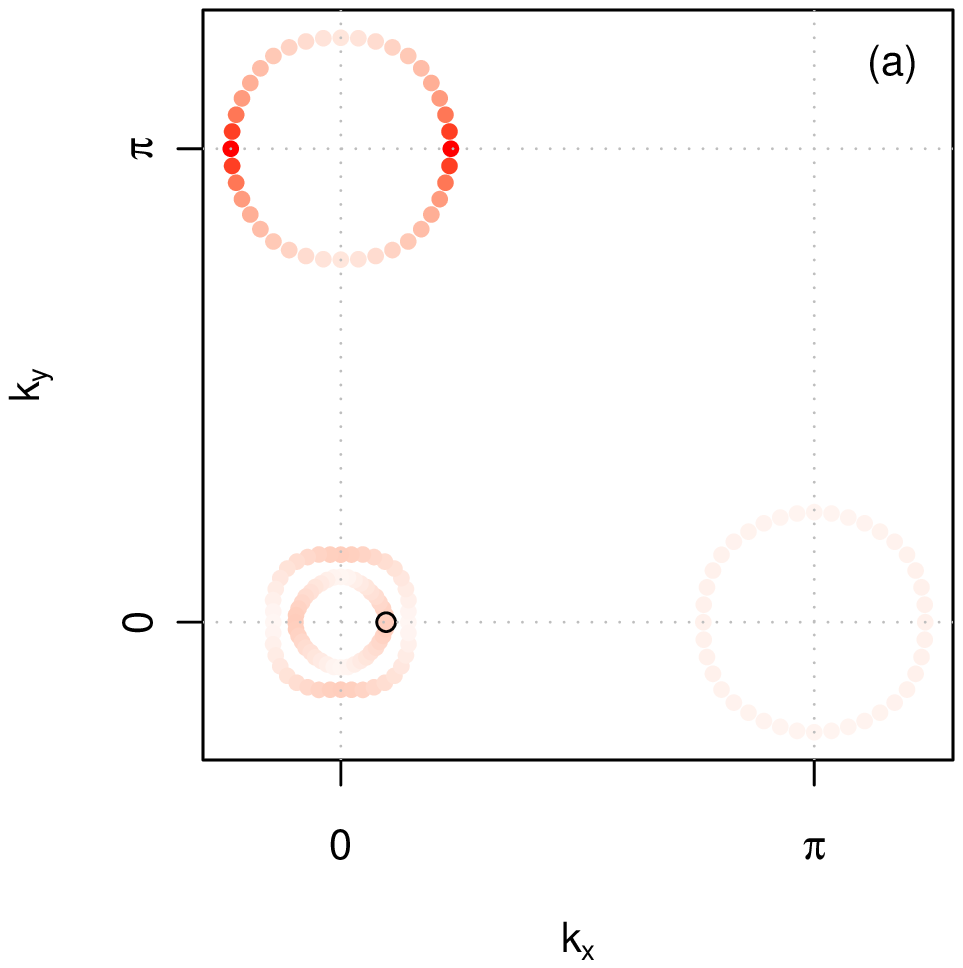} \vskip -1.5cm 
	\includegraphics[width=8cm,clip,angle=0]{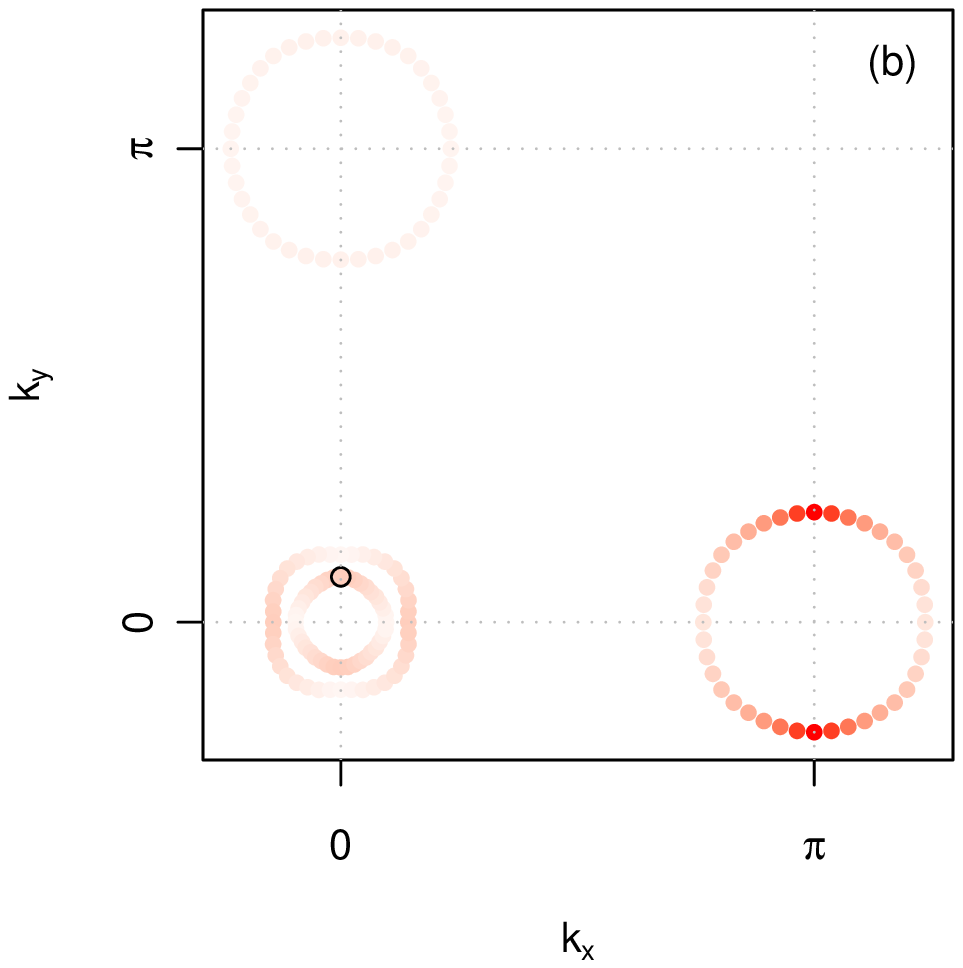} \caption{(color online) The strength $\Gamma_{ij}(k_0,k)$ associated with scattering a pair from the $\alpha_1$ Fermi surface with momenta a) $k_0=(k_F\hat\imath,-k_F\hat\imath)$ and b) $k_0=(k_F\hat\jmath,-k_F\hat\jmath)$ as indicated by the black circles as a function of momentum $k$. The orbital weight factors favor scattering from $d_{xz}$ to $d_{xz}$ and $d_{yz}$ to $d_{yz}$ orbital states.} \label{fig:3} 
\end{figure}

First we discuss the effect of the orbital matrix elements. The scattering strength $\Gamma_{ij}(k_0,k)$ for two different pairs on the $\alpha_1$-sheet with ${\bf k}_0=(k_F,0)$ and ${\bf k}_0=(0,k_F)$ as a function of momentum $k$ are illustrated in Fig.~\ref{fig:3}. 
For the former, Fig.~\ref{fig:3}a, the dominant scattering is to $d_{xz}(k\uparrow,-k\downarrow)$-pairs on the $\beta_2$ Fermi surface and for the latter, Fig.~\ref{fig:3}b, to $d_{yz}$-pairs on the $\beta_1$ Fermi surface. Note that the scattering strength is not simply a consequence of nesting; instead, it reflects the orbital weight structure factors. The $d_{xz}(k'\uparrow,-k'\downarrow)$ pairs on $\alpha_1$ scatter more strongly to $d_{xz}(k\uparrow,-k\downarrow)$ pair states on $\beta_2$, than to pair states on the $\beta_1$ Fermi surface which involve other orbitals. This means that while it is in general favorable to have a sign change between the gaps on the $\alpha_1$ and $\beta_2$ Fermi surfaces, the essential thing is to have a sign change of the gap between the red ($d_{xz}$) regions of the $\alpha_1$ Fermi surface and the red $(d_{xz})$ parts of the $\beta_2$ Fermi surface shown in Fig.~\ref{fig:1}. Likewise, one needs a sign change between the green ($d_{yz}$) regions of the $\alpha_1$ Fermi surfaces and the green ($d_{yz}$) $\beta_1$ Fermi surface. It is not important to maintain this sign change in the yellow ($d_{xy}$) regions of the $\beta$ Fermi surfaces. We will in fact see that the magnitude of the gap is larger on the $d_{yz}$ portion of the $\beta_1$ Fermi surface and smaller on the $d_{xy}$ parts. This actually leads to a small increase in the $\alpha$--$\beta_1$ pairing compared to the isotropic sign-switched gap.

In addition, as we will discuss, if there are other interactions such as the local Coulomb interaction or scattering processes such as inter-Fermi surface $\beta_1$--$\beta_2$ scattering, reducing the magnitude or even changing the sign of the gap on the $d_{xy}$ part of the $\beta$ Fermi surfaces can lead to an additional reduction of the short range Coulomb interaction and an enhancement of the pairing. In order to explore this latter effect, we need to obtain a more detailed accounting of the various contributions to the pairing strength. As discussed in Ref.~\onlinecite{graser:njp09}, for a given gap function $g(k)$, the effective pairing strength is determined from 
\begin{equation}
	\lambda[g(k)]=-\frac{\Sigma_{ij}\oint_{C_i}\frac{dk_\parallel}{v_F(k)}\oint_{C_j}\frac{dk'_\parallel}{v_F(k')}g(k) \Gamma_{ij}(k,k')g(k')}{(2\pi)^2\Sigma_i\oint_{C_i}\frac{dk_\parallel}{v_F(k)}[g(k)]^2} \label{eq:2} 
\end{equation}
Here $v_F(k)=|\nabla_kE_i(k)|$ for $k$ on a given Fermi surface $C_i$. In the following, we will normalize the gap function $g(k)$ such that the denominator is equal to the total one-electron density of states, 
\begin{equation}
	\sum_j\oint_{C_j}\frac{dk_{\parallel}g^2(k)}{2\pi\left(2\pi v_F(k)\right)}= \sum_jN_j(0)\,. \label{eq:3} 
\end{equation}
Then we can decompose $\lambda$ into its contributions from the different inter- and intra-Fermi surface scattering processes, 
\begin{equation}
	\lambda[g]=\sum_{ij}\lambda_{ij}[g] \label{eq:4} 
\end{equation}
with 
\begin{eqnarray}
	\lambda_{ij}[g] &=& -\oint_{c_i}\frac{dk_{\parallel}}{2\pi\left(2\pi v_F(k)\right)} \oint_{c_j}\frac{dk'_{\parallel}}{2\pi\left(2\pi v_F(k')\right)}\nonumber\\
	&&\times \frac{g(k)\Gamma_{ij}(k,k')g(k')}{\sum_iN_i(0)} \label{eq:5} 
\end{eqnarray}
\begin{figure}
	[htbp] 
	\vskip -1.5cm 
	\includegraphics[width=8cm,clip,angle=0]{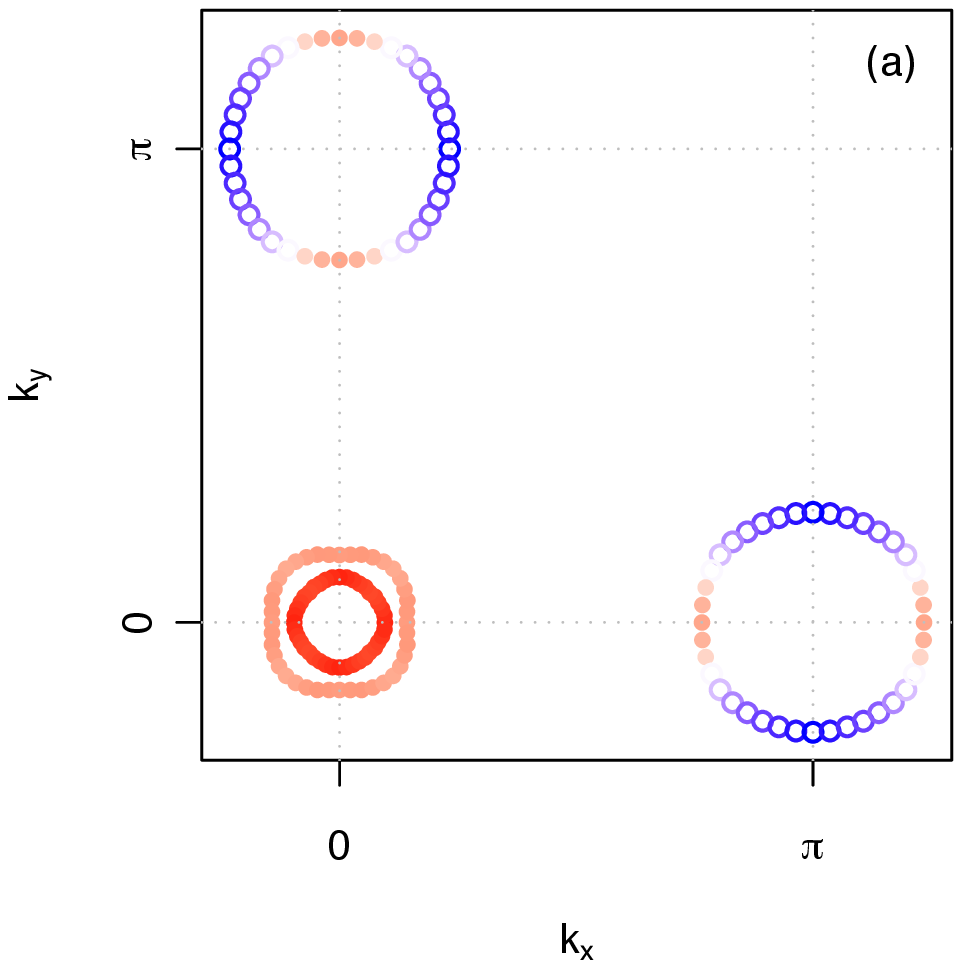} \vskip -1.5cm 
	\includegraphics[width=8cm,clip,angle=0]{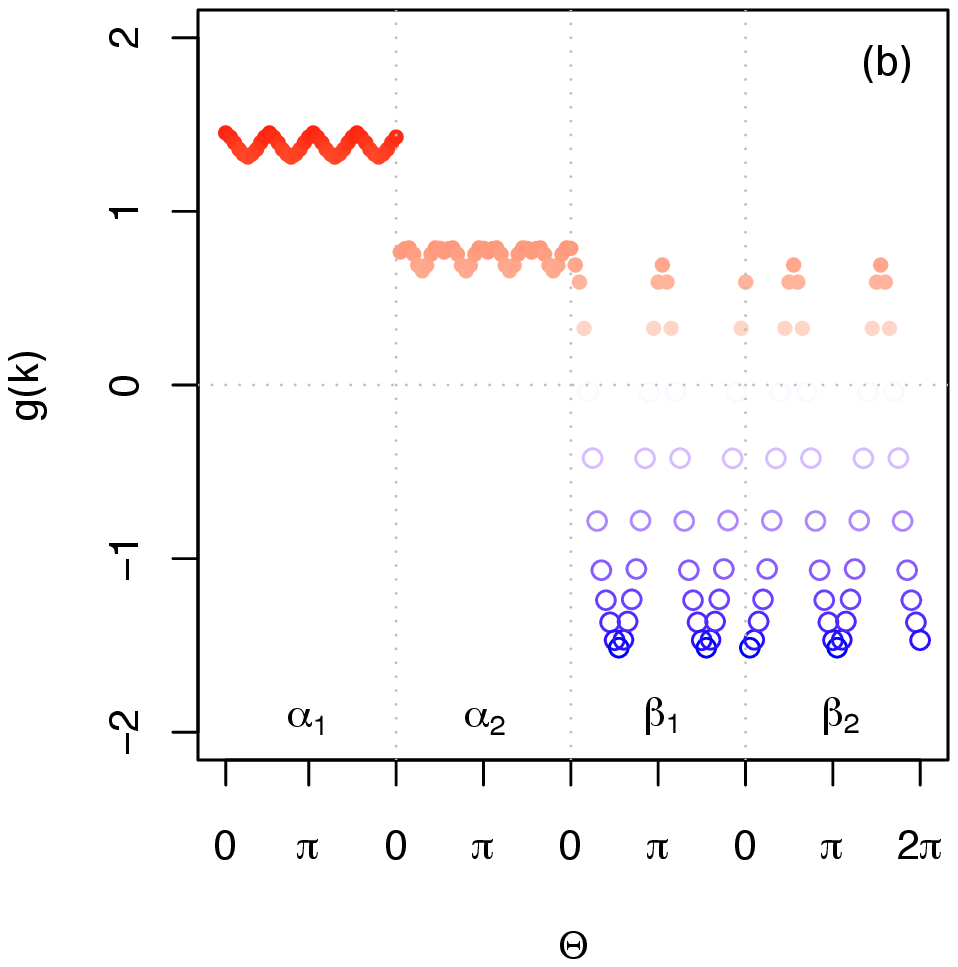} \vskip -1.0cm 
	\includegraphics[width=10cm,clip,angle=0]{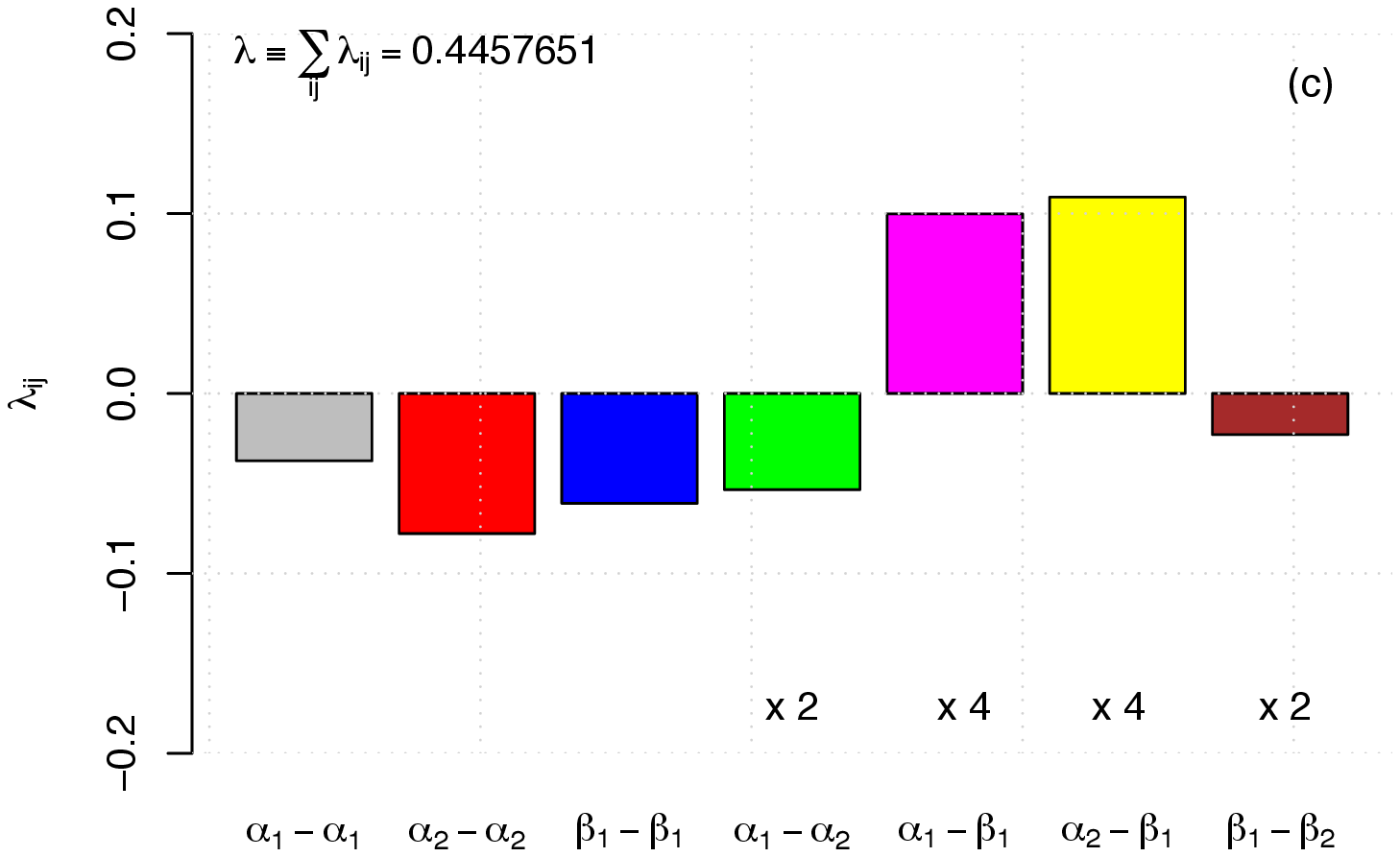} \caption{(color online) The RPA gap function a) plotted on the Fermi surfaces (red/solid circles positive and blue/open circles negative), (b) $g(k)$ as a function of angle and (c) bar graph of the various interaction components $\lambda_{ij}$. Note that the off-diagonal terms contribute to the total $\lambda$, Eq.~(\ref{eq:4}) with weights of two or four as indicated.} \label{fig:4} 
\end{figure}
First consider $\lambda_{ij}$ for two limiting cases (1) the optimal RPA $g(k)$ found as the variational solution of Eq.~\ref{eq:1a} and (2) a sign-switched $s$-wave with $g_{\alpha_1}=g_{\alpha_2}=1$ and $g_{\beta_1}=g_{\beta_2}=-1$. The results are shown in Figs.~\ref{fig:4} and \ref{fig:5}. 
\begin{figure}
	[htbp] 
	\vskip -1.5cm 
	\includegraphics[width=8cm,clip,angle=0]{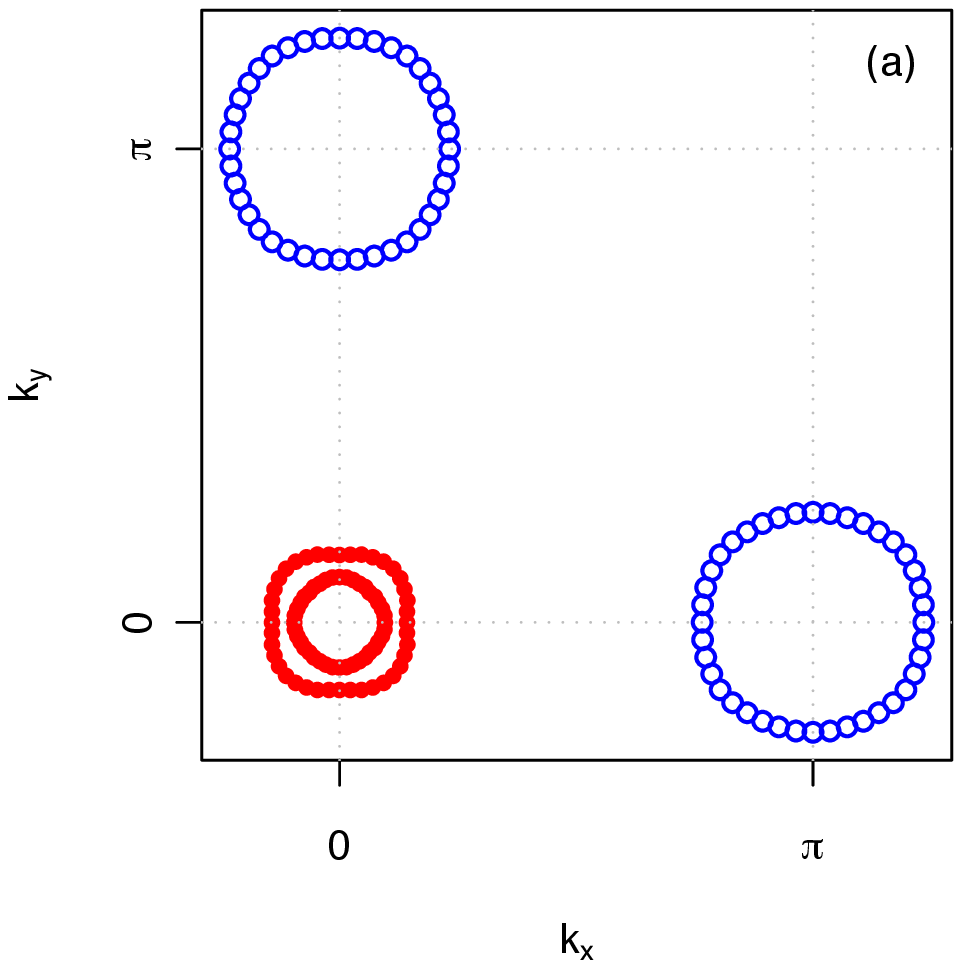} \vskip -1.5cm 
	\includegraphics[width=8cm,clip,angle=0]{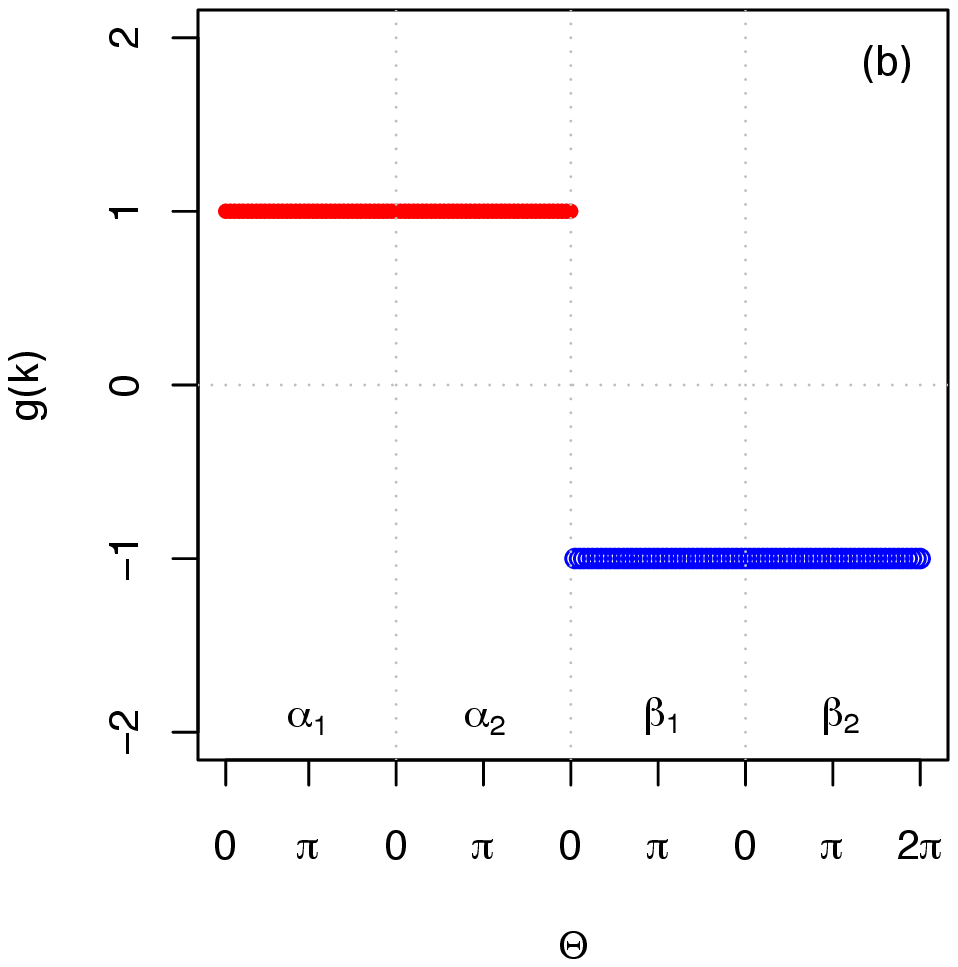} \vskip -1.0cm 
	\includegraphics[width=10cm,clip,angle=0]{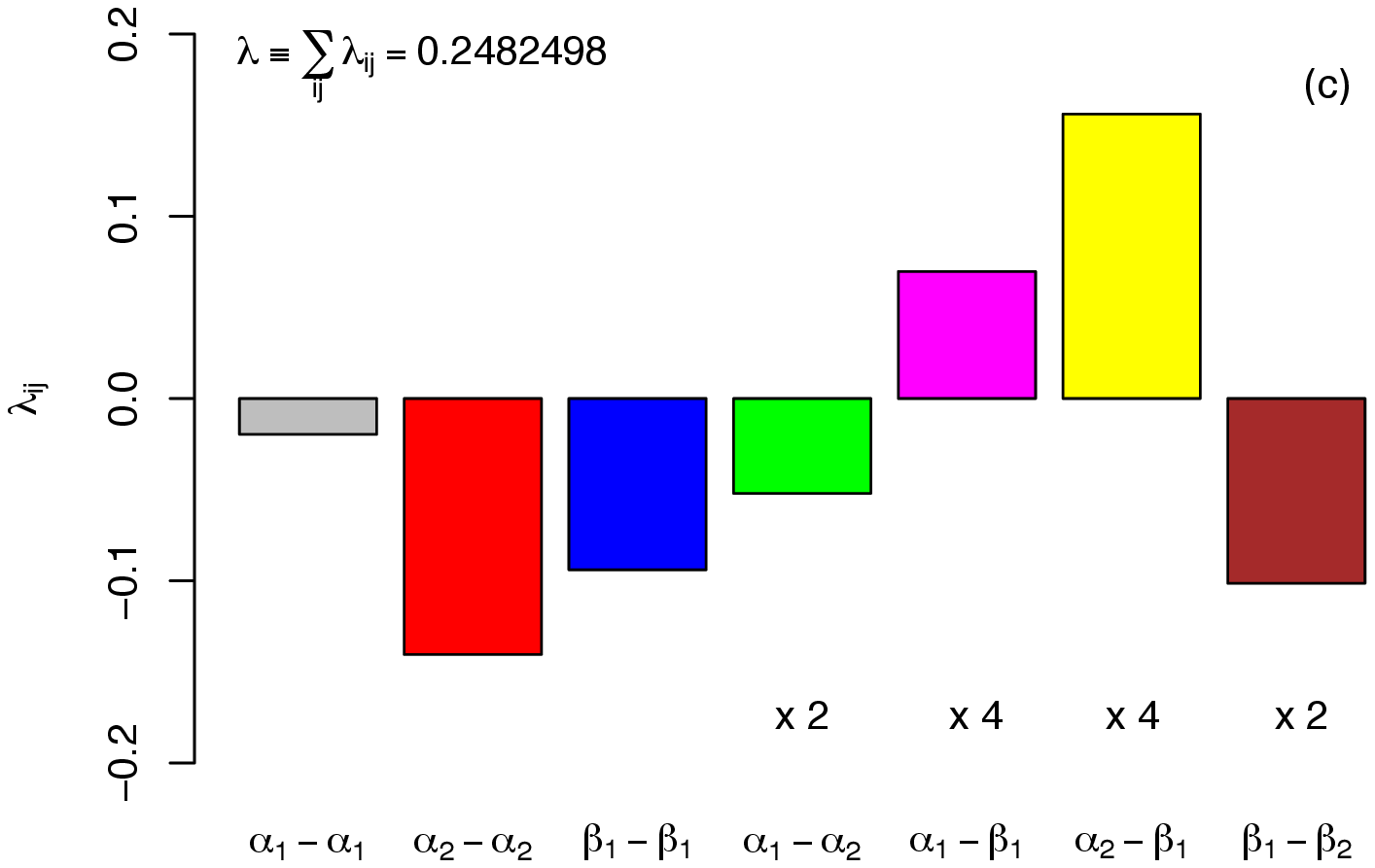} \caption{(color online) The sign-switched gap function (similar to Fig.~\protect\ref{fig:4}).} \label{fig:5} 
\end{figure}
The total pairing strength is significantly larger for the RPA $g(k)$. From the breakdown of the various contributions one sees that while the total $\alpha$--$\beta$ contribution is larger, 0.23 compared with 0.21, for the sign-switched gap function, the larger negative contribution of the Fermi surface $\lambda$'s as well as the larger negative $\beta_1$--$\beta_2$ contribution overcome this and $\lambda$ for the sign-switched gap is considerably smaller than the optimal $\lambda$. In Fig.~\ref{fig:4}, one can see that the anisotropy of the RPA solution is such that there are nodes on the $\beta$-sheets \footnote{We have also examined a parameter range in which a $d_{x^2-y^2}$ $B_{1g}$ gap is favored. Here the anisotropy occurs on the $\alpha_1$ and $\alpha_2$
Fermi surfaces where the gap has $d_{x^2-y^2}$ symmetry. This is dominantly driven
by the $\alpha-\beta$ scattering processes and arises naturally from the sign change
of the gap on the $\beta_1$ Fermi surface relative to the gap on the $\beta_2$
Fermi surface for the $B_{1g}$ state.}. As we will discuss, whether this happens or not depends upon the parameters. The interplay of the orbital weights in the pairing interaction, Eq.~\ref{eq:1a}, and the reduction of the intra $\beta_1$-$\beta_1$ Coulomb interaction, as well as the inter $\beta_1$-$\beta_2$ scattering lead to the anisotropy.

\begin{figure}
	[htbp] 
	\includegraphics[width=9cm,clip,angle=0]{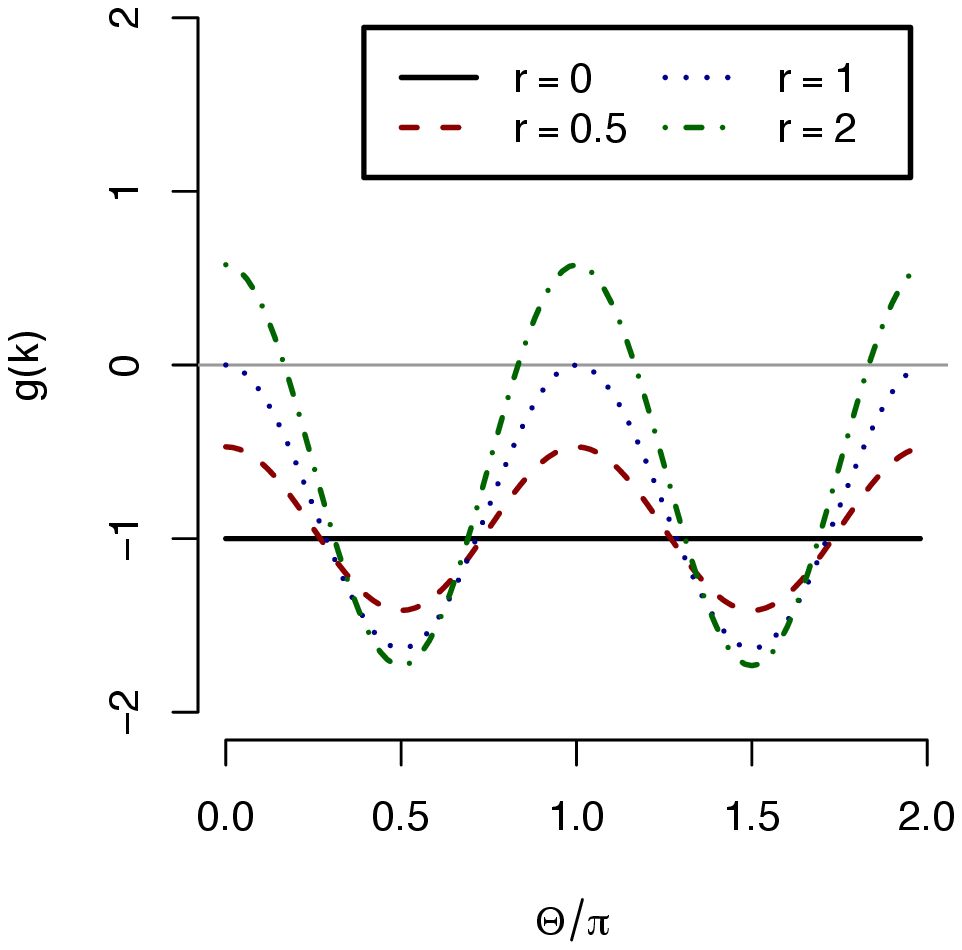}
		\caption{(color online) The phenomenological anisotropic gap function on the $\beta$ Fermi surface versus angle for several different values of $r$.} \label{fig:6} 
\end{figure}

\begin{figure}
	[htbp] 
	\includegraphics[width=9cm,clip,angle=0]{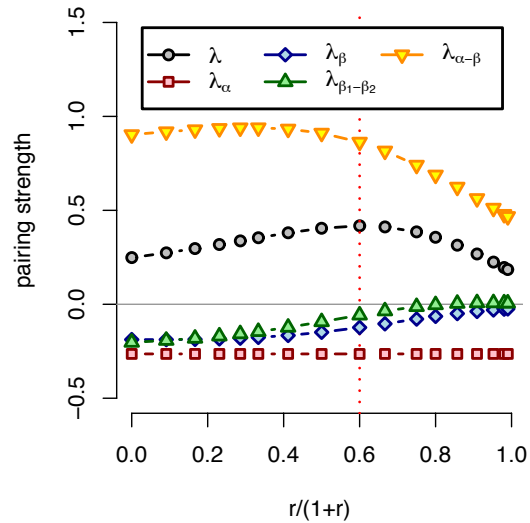}
	\caption{(color online) The pairing strength and its components versus $r$ for the anisotropic gap shown in Fig.~\protect\ref{fig:6}. Here $\lambda$ is the total pairing strength. The intra- and inter-$\alpha_1$ and $\alpha_2$ Fermi surface contributions are given by $\lambda_\alpha=\lambda_{\alpha_1\alpha_1}+\lambda_{\alpha_2\alpha_2}+ 2\lambda_{\alpha_1\alpha_2}$ and the intra-$\beta$ Fermi surface contribution by $\lambda_\beta=\lambda_{\beta_1\beta_1}+\lambda_{\beta_2\beta_2}$. The contribution to the pairing comes from $\lambda_{\alpha-\beta}=4\lambda_{\alpha_1\beta_1}+2\lambda_{\alpha_2\beta_1}$, while for this $A_{1g}$ gap the inter-$\beta_1-\beta_2$ contribution opposes the pairing for small $r$ and is neutralized for a large anisotropy. The red dashed line indicates the optimal degree of anisotropy, $r=1.5$.} \label{fig:7} 
\end{figure}

To illustrate this, consider the simple parameterization of an anisotropic gap with $g_{\alpha_1}=g_{\alpha_2}=1$ and 
\begin{equation}
	g_\beta=-a(1-r\cos2\theta) \label{eq:8} 
\end{equation}
Here $a=\left(2/(2+r^2)\right)^{1/2}$ so that the normalization condition, Eq.~(\ref{eq:3}) is satisfied. When $r=0$, $g_\beta=-1$ and we have the sign-reversed state previously discussed. Then as $r$ increases, the gap function becomes anisotropic on the $\beta$ Fermi surfaces as shown in Fig.~\ref{fig:6}. In Fig.~\ref{fig:7} we plot the total pairing strength $\lambda$ and some of its components versus $r$. 
Initially, as $r$ increases, the gap on the $\beta$ Fermi surface sheets becomes anisotropic, and the total pairing strength $\lambda$ increases. The slight increase in $\lambda_{\alpha-\beta}$ reflects the increase in the amplitude of the gap in the regions where the $d_{xz}$ and $d_{yz}$ orbital weights are largest. The additional increase in $\lambda$ arises from the suppression of the intra- and inter-Coulomb repulsion associated with the $\beta_1$ and $\beta_2$ Fermi surfaces due to the gap anisotropy. Finally, as $r$ increases further, the reduction of the $\alpha$--$\beta$ pairing contribution due to the anisotropy becomes larger than the suppression of the Coulomb interaction and the total pairing strength $\lambda$ decreases. For the interaction parameter set that we are using, this occurs for $r\simeq 1.5$, so that there are well developed nodes on the $\beta$-Fermi surfaces. However, if the intra- and inter-scattering on the $\beta$-Fermi surfaces were reduced, the nodes could be lifted. 

\section*{Conclusion}

The anisotropy of the $A_{1g}$ gap on the electron Fermi surfaces found in RPA and numerical functional renormalization group studies has been shown to arise from an interplay of three sources: (1) the variation of the weighting of the different $d$-orbitals on the Fermi surfaces, (2) the need to suppress the Coulomb repulsion and (3) the need to reduce the effects of the repulsive scattering between the electron-$\beta$ sheets. The orbital weight variation is familiar in other multi-orbital superconductors such as MgB$_2$ \cite{choi:nat02}. In essence, the pairing interaction is strongest between fermions in near neighbor $d_{yz}$ orbitals along the $x$-direction and near-neighbor $d_{xz}$ orbitals along the y-direction. The pairing associated with the $d_{xy}$ near neighbor orbitals is weaker. At the same time, the anisotropy leads to a reduction of the repulsive Coulomb interactions and the inter-Fermi surface $\beta_1$-$\beta_2$ scattering. As discussed, there is a balance between these effects which determine whether the anisotropy is sufficiently large that there are nodes on the $\beta$-sheets. The fact that the dominant orbital weights on the $\alpha_1$ and $\alpha_2$ Fermi surfaces are associated with the $d_{xz}$ and $d_{yz}$ orbitals leads to a more isotropic gap on these sheets.

The results that we have discussed are based upon a weak coupling spin-fluctuation approach. In particular, the various calculations have used an RPA form for the pairing vertex. Here the bandstructure and filling, along with the relative strengths of the onsite Coulomb and exchange interactions enter. A key feature of this approach is that the gap exhibits large anisotropies and that nodes may appear on the $\beta$-Fermi surfaces in the $s$-wave $A_{1g}$ state. However, one can imagine that variations in the parameters can alter the degree of anisotropy. 
An important implication of this is that it may provide an explanation for the wide variety of experimental indications that nodes are present in some cases and not in others. Another explanation for these apparent discrepancies could be the degree of disorder in different samples as discussed in Refs.~\onlinecite{mishra:prb09,vorontsov:arxiv09,parker:prb09,chubukov:prb08}.

\section*{Acknowledgements} We would like to acknowledge useful discussions with A. Chubukov. TAM and DJS would like to acknowledge support from Oak Ridge National Laboratory's Center for Nanophase Materials Sciences and the Scientific User Facilities Division, Office of Basic Energy Sciences, U.S.~Department of Energy. PJH would like to acknowledge support from the DOE grant DOE DE-FG02-05ER46236. DJS also thanks the Stanford Institute of Theoretical Physics for its hospitality. 

\bibliography{mybib.bib}

\end{document}